# Strain-released epitaxy of GaN enabled by compliant single-crystalline metal foils

Yaqing Ma[1, 2, 9], Junwei Cao[1, 2, 9], Huaze Zhu[2, 9]*, Yijian Song[4, 9], Huicong Chen[7], Menglin He[3], Jun Yang[5], Ping Jiang[5], Tong Jiang[2], Han Chen[1, 2], Xiang Xu[1, 2], Yuqiao Zheng[1], Hao Wang[2], Muhong Wu[6], Yu Zou[7], Xiaochuan Chen[8]*, Tongbo Wei[4]*, Kaihui Liu[6]*, Wei Kong[2]*.

[1] School of Materials Science and Engineering, Zhejiang University, Hangzhou 310027, China

[2] Department of Material Science and Engineering, School of Engineering, Westlake University, Hangzhou 310000, China

[3] Zhongke Crystal Materials (Dongguan) Technology Co., Ltd., Dongguan 523000, China

[4] Research and Development Center for Semiconductor Lighting Technology Institute of Semiconductors, Chinese Academy of Sciences, Beijing 100083, China

[5] Westlake Smoky-mountain Technologies, Hangzhou 310000, China

[6] State Key Laboratory for Mesoscopic Physics, Frontiers Science Centre for Nano-optoelectronics, School of Physics, Peking University, Beijing

[7] Department of Materials Science and Engineering, University of Toronto, Toronto, ON, M5S 3E4, Canada

[8] BOE Technology Group Co., Ltd., Beijing 100176, China

[9] These authors contribute equally: Yaqing Ma, Junwei Cao, Huaze Zhu, Yijian Song.

*Email:

chenxiaochuan@boe.com.cn; tbwei@semi.ac.cn; khliu@pku.edu.cn; kongwei@westlake.edu.cn.

# Abstract

Heteroepitaxy conventionally relies on rigid crystalline substrates, implicitly assuming that lattice and thermal mismatch must be accommodated within the epitaxial layer, leading to residual strain and defects that worsen with increasing substrate size. Here we demonstrate a substrate-mediated strain-partitioning regime in which lattice and thermal mismatch are preferentially partitioned into the substrate rather than stored in the epitaxial layer. We report the epitaxial growth of single-crystalline GaN on mechanically compliant yet crystallographically ordered single-crystalline copper foils. Atomic-resolution microscopy, geometric phase analysis and density functional theory reveal that mismatch-induced stress is primarily screened by elastic deformation of the Cu lattice, accompanied by localized interfacial slip confined to a few atomic layers, leaving the AlN and GaN epilayers nearly strain-free despite large nominal mismatch. Leveraging this strain-released epitaxial platform, we further demonstrate dense GaN micro-light-emitting diode arrays that benefit from efficient vertical electrical conduction and thermal dissipation enabled by the metallic substrate. By establishing compliant single-crystal metal foils as a new substrate class, this work identifies mechanical contrast as an underexplored governing parameter in heteroepitaxial design, with implications extending beyond GaN.

# Introduction

Heteroepitaxy underpins a wide range of modern electronic and optoelectronic materials, yet its foundational assumptions have remained largely unchanged for decades. In most epitaxial systems, the substrate is implicitly treated as a rigid template, such that lattice and thermal mismatch are accommodated predominantly within the epitaxial layer itself [1,2]. This paradigm inevitably leads to residual strain [3], misfit dislocations [4-6], and wafer bowing [3], particularly in systems with large lattice or thermal expansion mismatch [7,8]. As substrate dimensions increase, these strain-related effects intensify, imposing fundamental limits on epitaxial uniformity, defect control, and scalable integration [8].

Gallium nitride (GaN) epitaxy exemplifies these challenges. Despite its central role in light-emitting diodes [9,10], laser diodes [11], and high-power electronics [12], GaN is typically grown on rigid substrates such as sapphire, silicon, or silicon carbide, all of which exhibit substantial lattice and thermal mismatch with GaN [7,8,13]. Consequently, strain accumulation, dislocation formation, and wafer bowing remain intrinsic to large-area GaN heteroepitaxy, particularly as wafer sizes approach industrially relevant dimensions [8]. To mitigate mismatch-induced strain, a wide range of strategies have been explored, including buffer layer engineering [14,15], geometry-mediated relaxation [16], released substrates and membrane-based approaches [17-20], as well as van der Waals epitaxy [21,22]. These approaches have enabled incremental improvements in crystalline quality and strain relaxation, and have supported important advances in GaN materials and devices.

Despite their diversity, however, these strategies share a common limitation: strain is either confined within the epitaxial layer or avoided altogether, rather than actively accommodated by the substrate during epitaxial growth. In buffer-layer schemes, strain relaxation is primarily achieved through elastic distortion, compositional grading, or misfit dislocation formation within epitaxial layers, while the underlying substrate remains mechanically rigid. In geometry-mediated relaxation approaches, geometrical confinement and free surfaces enable macroscopic deformation of the substrate, such that strain relaxation is governed by geometry-enabled stress dilation. Released substrate and membrane-based strategies enable strain relaxation mainly after growth, through removal of the rigid substrate to release the stored elastic energy. Van der Waals epitaxy circumvents lattice mismatch through weak interfacial bonding, effectively decoupling strain but precluding strong epitaxial registry.

A fundamentally different approach is to reconsider the mechanical role of the substrate itself. If the substrate were both crystallographically ordered and mechanically compliant, mismatch strain would no longer need to be confined to the epitaxial layer but could instead be redistributed across the heterointerface into the substrate. Such a strain-partitioning mechanism, operating during epitaxial growth, requires a materials platform that combines long-range single crystallinity with sufficient mechanical compliance to allow localized elastic deformation or interfacial slip in the substrate, rather than strain storage or dislocation formation within the epitaxial layer. Conventional single-crystal wafers, including oxides and covalently bonded semiconductors, lack this combination of properties and are therefore intrinsically unsuitable for realizing such a regime. Separately, prior demonstrations of GaN or aluminum nitride (AlN) growth on bulk single-crystal metal substrates have focused primarily on crystallographic alignment, interfacial chemistry, or film morphology, without invoking or experimentally resolving a substrate-mediated strain accommodation mechanism at the atomic level [23-27].

Here we demonstrate that single-crystalline metal foils provide a new materials platform for heteroepitaxy, enabling a distinct strain-accommodation regime governed by mechanical contrast across the interface. Using single-crystalline copper (Cu) foils that are manufacturable at large scale via roll-to-roll crystallization, we realize high-quality GaN heteroepitaxy in which lattice and thermal mismatch are predominantly partitioned into the substrate rather than stored in the nitride epilayers. Atomic-resolution scanning transmission electron microscopy, geometric phase analysis (GPA), and density functional theory (DFT) calculations reveal that mismatch-induced stress is relieved through localized interfacial slip and elastic deformation within the Cu lattice, leaving the AlN and GaN layers nearly strain-free despite the large nominal lattice mismatch.

By establishing mechanically compliant single-crystal metal foils as a new substrate class for heteroepitaxy, this work identifies mechanical contrast as an underexplored governing parameter in strongly bonded heteroepitaxial systems, complementing conventional considerations of lattice symmetry and mismatch. While GaN serves as a model system, the demonstrated strain-partitioning mechanism is expected to be broadly applicable to other stiff-on-soft epitaxial combinations. To validate the functional implications of this platform, we further demonstrate dense GaN micro-light-emitting diode (LED) arrays that leverage the intrinsic electrical and thermal advantages of the metallic substrate.

## Single-crystalline GaN epitaxy on Cu foil

To establish compliant single-crystalline metal foils as a new epitaxial materials platform, we first examine whether long-range crystallographic order and mechanical compliance can be simultaneously achieved in a scalable substrate suitable for heteroepitaxy. Cu foils are particularly attractive in this context because they combine high ductility with well-defined crystallography when properly engineered. Recent advances in metal crystallization have demonstrated that industrial polycrystalline Cu foils can be transformed into single-crystalline foils with uniform orientation over macroscopic areas via high-temperature (HT) annealing [28,29], providing a pathway toward large-area, low-cost single-crystal metal substrates.

Figure 1a outlines the materials strategy employed in this work. Industrial poly-crystalline Cu foils were converted into single-crystalline Cu (111) foils through thermal crystallization, yielding long-range orientational order across tens-of-centimeters-scale areas (see Fig. 1b and Fig. S1). Electron backscatter diffraction (EBSD) inverse pole figure maps confirm uniform crystallographic alignment both out-of-plane and in-plane (see Fig. 1e), establishing a prerequisite for unidirectional epitaxial growth.

Unlike conventional bulk single-crystal metal substrates, the resulting Cu foils are only tens of micrometers thick and thus mechanically compliant, enabling local elastic deformation under epitaxial stress. A central challenge in utilizing such compliant foils for epitaxy is maintaining mechanical stability at the elevated temperatures required for nitride growth. To address this, the single-crystalline Cu foils were laminated onto a tungsten (W) carrier substrate via a metal thermo-compression bonding process (see Fig. 1a, c and Fig. S2). The W support provides mechanical rigidity and dimensional stability, while preserving the interfacial compliance of the Cu layer owing to its relatively low elastic modulus and yield strength. Importantly, W exhibits a thermal expansion coefficient closely matched to that of GaN, minimizing macroscopic thermal stress during growth without suppressing local strain accommodation within the Cu foil.

Using this composite compliant substrate, GaN epitaxy was carried out via hydride vapor phase epitaxy (HVPE), a mature, high-throughput technique widely used for producing large-area, high-quality GaN epitaxial materials [30]. The growth sequence begins with an AlN seed layer followed by a low-temperature (LT) GaN layer (see Fig. 1a, Fig. S3 and Methods), analogous to conventional GaN-on-sapphire processes (see Fig. S4) [31]. The resulting GaN films exhibit highly ordered surface morphology (see Fig. 1d and Fig. S5) and crystallography: Z- and Y-direction inverse polar figure (IPF) maps show excellent in-plane and out-of-plane crystalline alignment (see Fig. 1f), and atomic force microscopy (AFM) measurements reveal atomically smooth step-flow terraces with a root-mean-square (RMS) roughness of 0.59 nm (see Fig. 1g), comparable to state-of-the-art epitaxial GaN.

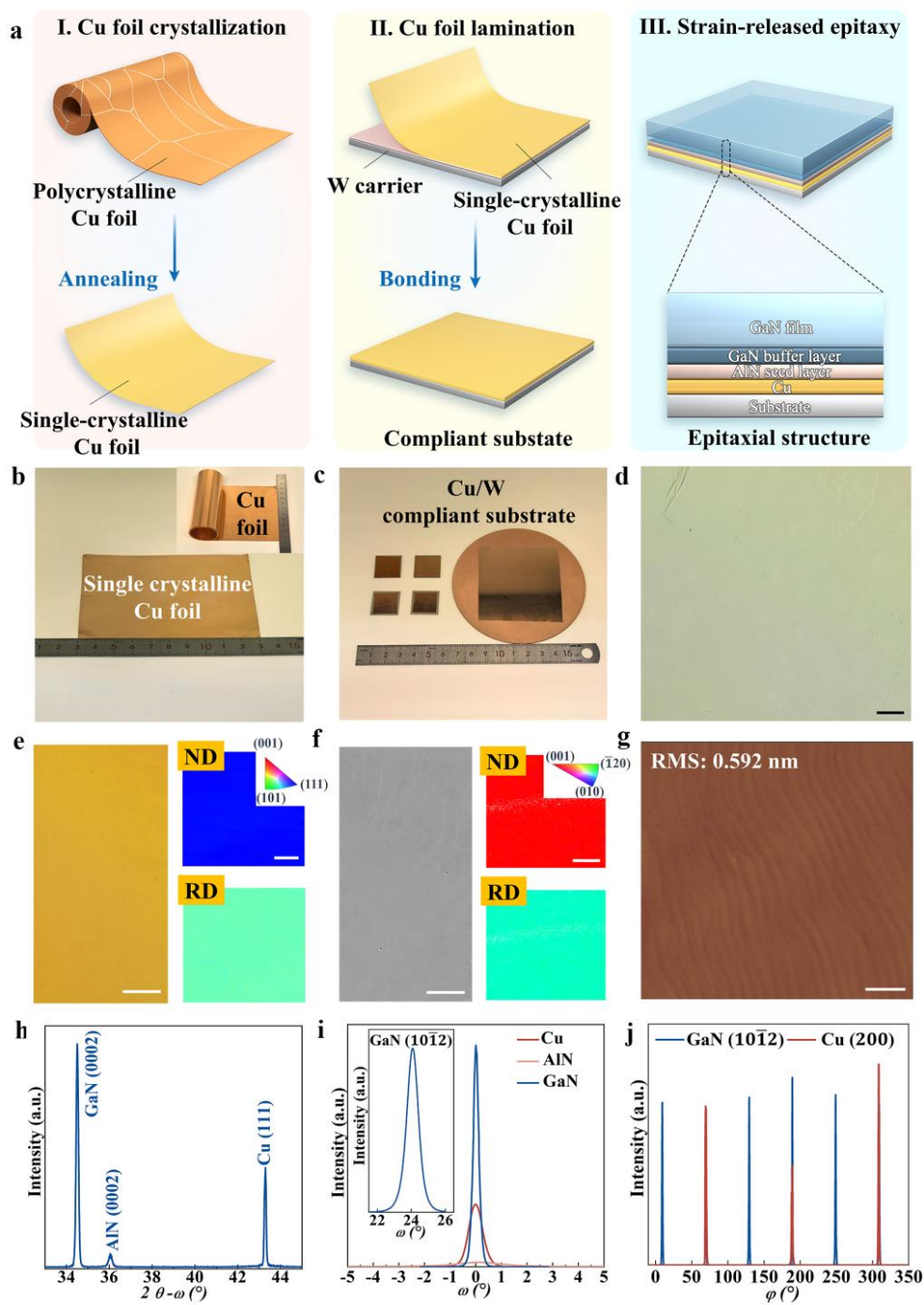

**Fig. 1. Fabrication strategy and structural characterization of GaN epitaxy on compliant single-crystalline Cu foils. a**, Schematic illustration of the epitaxial platform, showing conversion of poly-crystalline Cu foils into single-crystalline Cu (111), subsequent bonding to a W carrier, and GaN heteroepitaxy. **b**, Photograph of a Cu foil after high-temperature crystallization, demonstrating large-area single crystallinity. **c,** Photograph of bonded Cu foils on a W carrier, illustrating mechanical integrity and scalability of the composite substrate. **d**, Optical micrograph of GaN grown on Cu foil, showing uniform surface morphology. Scale bar:

50 µm. **e**, Optical micrograph (scale bar: 20 µm) and corresponding EBSD IPF maps (scale bar: 200 µm) of the Cu foil along the normal and rolling directions, confirming long-range single crystallinity. **f**, Plan-view SEM image (scale bar: 20 µm) of the GaN surface and EBSD IPF maps (scale bar: 200 µm) along the Z- and Y-directions, revealing excellent in-plane and out-of-plane crystalline alignment. **g**, AFM image of GaN grown on Cu foil, showing atomically smooth step-flow terraces with an RMS roughness of 0.59 nm. Scale bar: 0.5 nm. **h**, XRD $\omega$-$2\theta$ scan of the GaN/AlN/Cu layer stack, exhibiting sharp GaN (0002), AlN (0002) and Cu (111) reflections. **i**, Rocking curves of GaN (0002), AlN (0002), and Cu (111); the inset shows the rocking curve for GaN ($10\bar{1}2$). **j**, In-plane $\varphi$ scan of GaN (blue line) and Cu (red line) reflections, confirming single-crystalline GaN and Cu foil, and their in-plane alignment.

The epitaxial relationships among GaN, AlN, and Cu are confirmed by X-ray diffraction. $\omega$-$2\theta$, and $\omega$ scans show distinct GaN (0002), AlN (0002), and Cu (111) reflections (see Fig. 1h and 1i), indicating the absence of polycrystalline grains. $\varphi$ scans of GaN exhibit strong six-fold symmetry along with Cu (200) peaks (see Fig. 1j), demonstrating excellent in-plane alignment between the epitaxial GaN and single-crystalline Cu lattice.

These observations demonstrate that mechanically compliant single-crystalline metal foils can simultaneously satisfy the crystallographic requirements for epitaxy and the mechanical conditions to sustain high-temperature processes.

## Material analysis of GaN epitaxy on single-crystalline Cu foil

To examine the epitaxial material quality of GaN on single-crystalline Cu foil, and elucidate the defect evolution, we performed a systematic comparison of GaN films grown on single-crystal Cu foil and on conventional c-plane sapphire using identical buffer stacks and growth conditions. In both cases, the process incorporated a 200-nm LT AlN layer and a 2-µm LT-GaN buffer, followed by a HT GaN layer of 3 µm. Cross-sectional High-Angle Annular Dark-Field (HAADF) imaging of the GaN/AlN/Cu structure reveal a uniform GaN film with threading dislocation density (TDD) at the order of $10^8$ cm$^{-2}$ (see Fig. S6). Cross-sectional TEM images show that dislocations are partially terminated at the GaN/AlN interface while some continue propagating to the top surface (see Fig. 2a). Electron-channeling contrast imaging (ECCI) further visualizes the corresponding dislocation at the top surface to be at the $5 \times 10^8$ cm$^{-2}$, corroborating the TDD extracted from cross-sectional TEM (see Fig. 2b and Fig. S6). For comparison, GaN grown on sapphire under the same epitaxial conditions displays markedly higher defect densities. TEM (see Fig. 2e and Fig. S6) and ECCI (see Fig. 2f) indicate a TDD of $3 \times 10^9$ cm$^{-2}$, including both screw and edge components—a level consistent with typical GaN-on-sapphire heteroepitaxy [1].

The AlN buffer layer plays an essential role in enabling high-quality GaN epitaxy on Cu. Sputtered AlN on Cu forms a columnar single-crystalline structure (see Fig. 2c) with metal-face polarity (see Fig. S7), whereas AlN grown on sapphire exhibits smaller grains and poorer columnar alignment (see Fig. 2g). Rocking-curve measurements of the AlN (0001) reflection show similar full width at half maximum (FWHM) values of ~1.6° for the two substrates (see Fig. 2d), yet the in-plane scan reveals substantially improved crystallinity for AlN on Cu (see Fig. S8). XRD further indicates that AlN on Cu is significantly more relaxed than AlN on sapphire: the AlN (002) peak shifts downward from the stress-free position at 36.01°, corresponding to a tensile strain of only 0.11% for AlN-Cu but 0.25% for AlN–sapphire. This behavior is counterintuitive. Despite the larger nominal lattice mismatch between AlN and Cu (21.7%) compared with AlN and sapphire (13.2%), the AlN film on Cu relaxes more efficiently and exhibits higher crystalline quality—an outcome not anticipated prior to this study.

Most remarkably, GaN grown on AlN/single-crystal Cu reaches a nearly strain-free state, with the residual strain reduced from 0.29% in GaN-on-sapphire to only 0.05% in GaN-on-Cu (see Fig. 2h). This dramatic reduction in strain indicates that the Cu foil induces far less interfacial stress during epitaxy than conventional sapphire substrates. The resulting relaxation directly contributes to the lower dislocation density and mitigates common issues such as wafer bowing and warping that arise from strain accumulation in GaN-on-sapphire systems. Together, these results reveal that single-crystal Cu foil not only supports high-quality GaN epitaxy but also fundamentally alters the stress–defect landscape compared with conventional substrates, enabling substantially lower strain and reduced dislocation propagation.

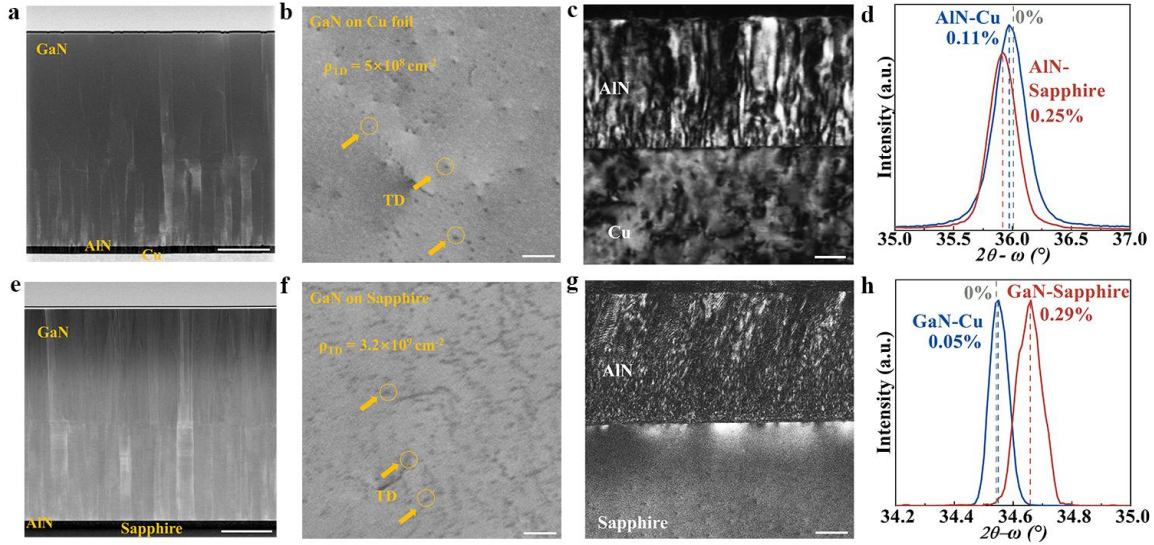

**Fig. 2. Dislocation characteristics and strain evolution of GaN grown on compliant Cu foils compared with rigid sapphire substrates. a**, Cross-sectional TEM image of GaN-on-Cu. Scale bar: 1 μm. **b**, ECCI of the GaN surface on Cu foil, confirming a TDD of ~5 × $10^8$ cm$^{-2}$. Scale bar: 200 nm. **c**, Dark-field TEM image of AlN (0001) film grown on Cu (111). Scale bar: 50 μm. **d**, The comparison of $2\theta$-$\omega$ curve of AlN (0001) grown on Cu foil (blue line) and sapphire (red line), respectively. **e**, Cross-sectional TEM image of GaN-on-sapphire. Scale bar: 1 μm. **f**, ECCI of GaN-on-sapphire, indicating a TDD of ~3 × $10^9$ cm$^{-2}$. Scale bar: 200 nm. **g**, Dark-field TEM image of AlN (0001) film grown on sapphire (0001), showing smaller grains and reduced columnar quality. Scale bar: 50 μm. **h**, The comparison of $2\theta$-$\omega$ curve of GaN (0001) grown on Cu foil (blue line) and sapphire (red line).

## Interfacial strain partitioning governed by mechanical contrast

To elucidate the physical origin of the strain relaxation observed in GaN grown on single-crystalline Cu foils, we examine the atomic-scale structure and strain distribution at the AlN/substrate interface. In conventional heteroepitaxy on rigid substrates, mismatch strain is typically relieved through the formation of misfit dislocations within the epitaxial layer, while the substrate remains elastically undeformed [1,2]. This classical picture implicitly assumes that the substrate is mechanically rigid relative to the epitaxial film.

We first consider AlN grown on c-plane sapphire as a reference system that exemplifies this rigid-substrate regime. As shown in figure 3a-c, AlN adopts the well-established in-plane epitaxial relationship with a 30° rotation that minimizes the effective lattice mismatch [32]. High-resolution Scanning Transmission Electron Microscope (STEM) images reveal a periodic array of misfit dislocations at the AlN/sapphire interface, consistent with classical misfit accommodation models [32]. GPA further shows that the resulting strain is concentrated almost entirely within the AlN layer, while the sapphire substrate remains essentially strain-free (see Fig. 3d). These observations confirm that, in the absence of substrate compliance, lattice mismatch is accommodated through elastic distortion and dislocation formation in the epitaxial film. In striking contrast, the AlN/Cu interface exhibits a fundamentally different strain accommodation behavior. AlN grows epitaxially on Cu (111) with a well-defined in-plane registry, despite the substantially larger nominal lattice mismatch. Based on lattice parameters, a coincidence lattice with a five-to-six matching periodicity is expected (see Fig. 3e), which would classically give rise to a dense network of misfit dislocations at the interface. However, atomic-resolution HAADF-STEM imaging reveals no discernible misfit dislocations or lattice distortion within the AlN layer (see Fig. 3f-g). Instead, the Cu atomic planes adjacent to the interface exhibit pronounced periodic shear and displacement, indicating that mismatch strain is redistributed into the metallic substrate rather than stored in the AlN film.

This unconventional strain accommodation is directly visualized by GPA strain mapping. As shown in figure 3h, the AlN layer near the interface remains nearly strain-free, whereas significant tensile strain develops within the Cu lattice over only a few atomic layers from the interface. The spatial localization of strain within Cu stands in sharp contrast to the sapphire case and demonstrates a reversal of the conventional strain-partitioning hierarchy in heteroepitaxy.

The origin of this behavior lies in the pronounced mechanical contrast across the AlN/Cu interface. Cu possesses an in-plane elastic modulus (~120 GPa) [33] substantially lower than that of AlN (~160 GPa) [34] and sapphire (~470 GPa) [35] as well as a yield strength (~70 MPa) [36] orders of magnitude smaller than that of covalently bonded AlN (~25 GPa) [34]. As a result, the interfacial mismatch stress generated during AlN nucleation exceeds the local yield threshold of Cu, triggering elastic deformation and atomic slip within the metal lattice. This deformation effectively screens the epitaxial layers from accumulating strain, thereby suppressing the formation of misfit dislocations in AlN.

To further substantiate this strain-partitioning mechanism, we performed DFT calculations using the experimentally observed AlN/Cu epitaxial registry (see Fig. 3e and Methods). Structural relaxation of the AlN/Cu interface reveals that atomic displacements are strongly localized on the Cu side, while the AlN lattice rapidly recovers its bulk-like structure away from the interface (see Fig. 3i, Fig. S9 and S10). The calculated in-plane strain distribution corroborates the experimental GPA results, showing that elastic strain is confined to the Cu lattice within a narrow interfacial region (see Fig. 3g-i). These calculations confirm that localized interfacial slips in Cu provide an energetically favorable pathway for mismatch accommodation.

Importantly, this mechanically driven strain partitioning persists beyond lattice mismatch to thermal mismatch. Despite the large difference in thermal expansion coefficients between GaN (~$4.5\times10^{-6}$ K$^{-1}$) and Cu (~$17\times10^{-6}$ K$^{-1}$) [37,38], finite element modeling reveal that thermal strain is likewise preferentially absorbed by the Cu foil (see Fig. S11), particularly when supported by a W carrier with a closely matched thermal expansion coefficient (~$4.8\times10^{-6}$ K$^{-1}$) [39]. As a consequence, the GaN epilayer experiences substantially reduced thermal stress upon cooling compared with GaN grown on sapphire, further contributing to the observed reduction in residual strain and threading dislocation density.

Together, these results establish a distinct heteroepitaxial regime in which strain accommodation is governed not by lattice matching and thermal matching alone (see Fig. 3j, k) [40-55], but also by mechanical contrast across the interface. In this regime, a compliant yet ordered substrate surface actively participates in strain relaxation, fundamentally altering the stress–defect landscape of heteroepitaxy.

In addition to its mechanical compliance, the single-crystal Cu foil platform offers a decisive scalability advantage. Unlike wafer-size–limited alternatives, single-crystalline Cu foils can be fabricated at meter-scale through roll-to-roll crystallization, combining low cost, high throughput, and long-range crystalline order (see Fig. 3l) [28,29]. This scalability positions single-crystal Cu foil as a highly promising new class of substrates for large-area, high-quality GaN optoelectronics.

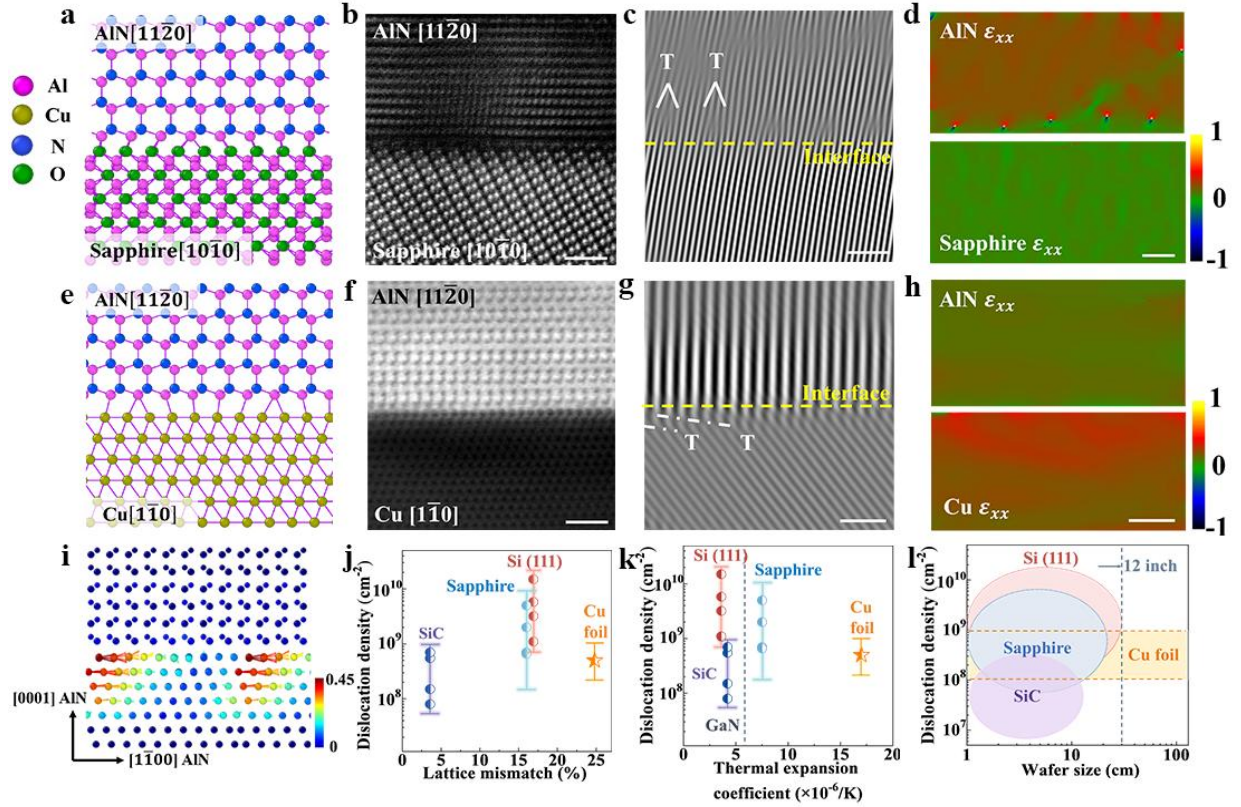

**Fig. 3 Atomic-scale visualization of strain partitioning governed by mechanical contrast at epitaxial interfaces. a**, DFT-constructed AlN/sapphire interface, representing the rigid-substrate case, where one misfit dislocation is expected every six AlN unit cells. **b**, HAADF-STEM image of the AlN/sapphire interface. Scale bar: 1 nm. **c**, Inverse fast Fourier transform image highlighting a periodic array of misfit dislocations at the AlN/sapphire interface, consistent with classical dislocation-mediated strain relaxation. Scale bar: 1 nm. **d**, In-plane strain ($\varepsilon_{xx}$) map obtained by GPA, revealing that mismatch strain is localized within the AlN layer while the sapphire substrate remains essentially strain-free. Scale bar: 1 nm. **e,** DFT-constructed AlN/Cu interface, showing one misfit dislocation is expected for every five AlN unit cells. **f,** HAADF-STEM image of the AlN/Cu interface. Scale bar: 1 nm. **g**, Inverse fast Fourier transform image of the AlN/Cu interface highlighting pronounced lattice shear and atomic displacement within the Cu lattice adjacent to the interface, indicative of interfacial slip. Scale bar: 1 nm. **h**, GPA in-plane strain ($\varepsilon_{xx}$) demonstrating that strain is concentrated within the Cu substrate over a few atomic layers, while the AlN layer remains nearly strain-free. Scale bar: 1 nm. **i**, DFT-relaxed atomic displacement field at the AlN/Cu interface, confirming elastic-dominated strain screening accompanied by localized interfacial slip in Cu. **j–l**, Comparative plot summarizing lattice mismatch, thermal mismatch, achievable substrate size, and corresponding threading dislocation density for GaN heteroepitaxy on compliant single-crystalline Cu foils versus conventional rigid substrates.

## High-density micro-LEDs fabrication and performance characterizations

The single-crystalline Cu foil composite substrate provides a synergistic combination of mechanical robustness and functional advantages for device integration. The mechanical ductility of Cu facilitates strain relaxation during epitaxy, while the planar W support offers excellent dimensional stability, enabling large-area, high-precision lithography required for high-density device arrays. Beyond these mechanical benefits, the inherently high electrical and thermal conductivities of metals introduce capabilities unattainable on sapphire or Si substrates—most notably, efficient vertical heat spreading and vertical current injection.

A key challenge, however, is that although GaN can conduct heat vertically into the metal, the heteroepitaxial AlN buffer layer remains electrically insulating and prevents vertical conduction [56]. To fully leverage the advantages of a metallic substrate, we developed a selective-area-epitaxy (SAE)–enabled Ti edge-contact architecture that circumvents the insulating AlN and directly connects the GaN n-type region to the underlying metal (see Fig. 4a). This design streamlines the materials-to-device workflow and forms a fully metallic vertical conduction pathway. In this process, a Ti layer and then a $SiO_2$ hard mask were first deposited onto the single-crystalline AlN/Cu stack. Periodic openings were patterned and etched by inductively coupled plasma (ICP) etching down to the AlN seed layer. Subsequent SAE from these micro-openings yielded single-crystal GaN arrays (see Methods, Fig. S12 and S13). The Ti layer forms a lateral edge contact to the n-GaN sidewalls, and the matched Fermi-level alignment ensures efficient electron injection. The Ti electrode then connects vertically to the Cu substrate through a via, thereby bypassing the insulating AlN layer and establishing a continuous low-resistance vertical current pathway from the GaN active region to the metal substrate (see Fig. 4a).

Using this architecture, we fabricated 5-μm-pitch micro-LED arrays on the single-crystalline Cu foil composite substrate. The complete vertical LED structure from bottom to top consists of: W support, single-crystalline Cu foil, AlN buffer, Ti n-contact, $SiO_2$ mask, n-GaN, InGaN/GaN multi-quantum wells (MQWs), p-GaN, and the indium tin oxide (ITO) p-contact (see Fig. 4a). The GaN crystals exhibit excellent single crystallinity (see Fig. 4b) and long-range periodic order across the entire array (see Fig. 4c). Elemental mapping and high-resolution STEM confirm sharp and well-defined MQW interfaces (see Fig. S14 and S15). The pyramidal micro-LEDs expose clean semi-polar $(10\bar{1}1)$ facets with a 62° inclination, which effectively suppresses the quantum-confined stark effect (QCSE) and enhances radiative recombination [57]. The SAE approach also eliminates the plasma-etch damage that typically introduces nonradiative recombination pathways at device sidewalls, especially at reduced diameter [57].

Consequently, the micro-LEDs grown on single-crystal Cu foil exhibit excellent optical performance. Continuous-wave photoluminescence (PL) measurements yield an internal quantum efficiency (IQE) of 61.7% (see Fig. 4e), surpassing the performance of GaN LEDs of similar dimensions grown on conventional substrates [58,59]. Electrical characterization through the transmission-line method (TLM) shows that the Ti/n-GaN contact achieves a specific contact resistivity of $1.6 \times 10^{-5}$ Ω·cm$^2$ (see Fig. 4f), on par to the best value reported with commercial LED technology and allowing efficient carrier injection [60]. The vertical p-to-n conduction enabled by the Ti/Cu pathway allows the devices to exhibit intense electroluminescence in both macro- and micro-scales (see Fig. 4d).

Direct contact between GaN and the metallic substrate also enables highly efficient vertical heat extraction into the metal heat sink, owing to the more than ten-fold higher thermal conductivity of Cu (380 - 400 W·(m·K)$^{-1}$) compared with sapphire (35 - 45 W·(m·K)$^{-1}$) [61,62].

Under identical electrical power, micro-LEDs on Cu exhibit substantially lower surface temperatures than those on sapphire. At a power density of 1,300 W·cm$^{-2}$, the temperature rises of devices on Cu ($\Delta T$ = 8.9 °C) is approximately half that of their sapphire counterparts ($\Delta T$ = 17.4 °C) (see Fig. 4g). The demonstrated power handling of 1,300 W·cm$^{-2}$, together with a current density of 200 A·cm$^{-2}$, meets the requirements of representative mainstream LED applications. This includes high-power and high-current uses such as automotive lighting and outdoor lighting. Owing to the high-density emitting elements, this application is also well suited for display-relevant scenarios ranging from flat-panel displays to augmented-reality (AR) systems. These results establish a scalable light-emitting device platform with excellent electro-optical-thermal performance (see Fig. 4h) [63-71].

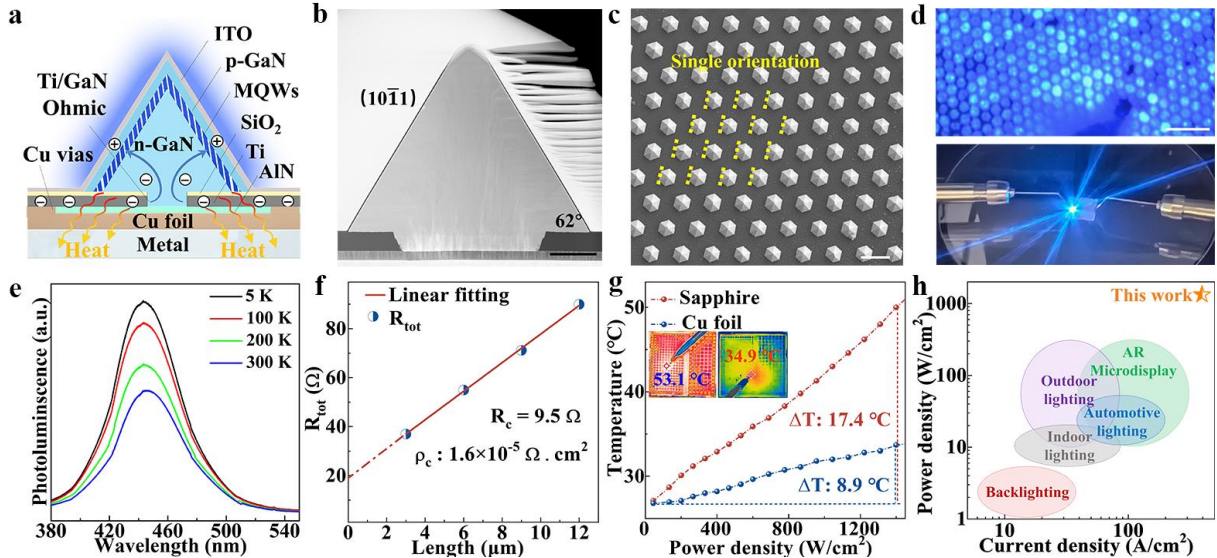

**Fig. 4. High-density GaN micro-light-emitting diode arrays enabled by the compliant metal epitaxial platform. a**, Schematic of the vertically conductive GaN micro-LED architecture, illustrating vertical thermal dissipation and the Ti edge-contact pathway that bypasses the insulating AlN buffer and enables vertical current injection into the micro-LED from metal substrate. **b**, STEM image of a pyramid-shaped GaN micro-LED grown on Cu foil, showing single crystallinity. Scale bar: 5 μm. **c**, SEM image of GaN pyramid arrays with a pitch of 5 μm, showing long-range single orientation alignment. Scale bar: 1 μm. **d**, Microscopic (top, scale bar: 20 μm) and macroscopic (bottom) optical images of the micro-LED devices in operation. **e**, Temperature-dependent PL spectra the InGaN/GaN MQWs. **f**, TLM characterization of the Ti/n-GaN contact, indicating low specific contact resistivity. **g**, Comparison of surface temperature rise for micro-LEDs on Cu foil and sapphire under identical power densities, highlighting enhanced vertical heat dissipation enabled by the metallic substrate. **h**, Comparison of power-density and current-density operating characteristics across representative GaN LED applications.

## Conclusions

In conclusion, we establish mechanically compliant single-crystalline metal foils as a new substrate class for heteroepitaxy, enabling a strain-accommodation regime fundamentally distinct from that of conventional rigid wafers. Using GaN as a model system, we demonstrate that lattice and thermal mismatch need not be confined to the epitaxial layer but can instead be predominantly partitioned into the substrate when crystallographic order and mechanical compliance coexist at the interface.

Atomic-scale characterization and first-principles calculations reveal that mismatch-induced stress is primarily screened through elastic deformation of the Cu lattice, accompanied by localized interfacial slip confined to a few atomic layers. This elastic-dominated strain screening suppresses residual strain accumulation and dislocation formation in the AlN and GaN epilayers despite large nominal mismatch, thereby redefining how strain and defects can be controlled in strongly bonded heteroepitaxial systems. The metallic substrate further enables vertical electrical conduction and efficient heat dissipation in GaN micro-scale light-emitting devices, providing functional validation of the strain-released epitaxial platform.

Beyond elucidating this interfacial strain-partitioning mechanism, the use of scalable single-crystalline metal foils removes fundamental size constraints imposed by rigid substrates and introduces mechanical contrast as a governing design parameter alongside lattice symmetry and mismatch. While demonstrated here for GaN, the underlying principle is expected to be broadly applicable to other stiff-on-soft epitaxial combinations, offering a general strategy for engineering the strain-defect landscape in lattice-mismatched materials systems.

# Methods

### Template preparation

Cu foils (1 cm$^2$, 2-inch, 4-inch) were first annealed at 1050 °C-1070 °C for 16 hours under a mixed hydrogen and argon atmosphere for an extended period to obtain single-crystalline Cu foils. The single-crystal Cu foils were then cut and transferred onto sapphire, glass, and metal substrates of matching sizes by Cu film bonding process. Subsequently, a single-crystalline AlN seed layer was deposited on the bonded Cu foils by PVD at 500 °C, using an Ar/N$_2$ mixed atmosphere and a sputtering power of 150 W.

### Epitaxy of GaN films using HVPE

GaN films were fabricated using a HVPE system. A two-step growth was employed: an 850 °C buffer and a 1000 °C HT layer. Firstly, high purity 7N Ga metal was placed in a pre-reaction chamber and reacted with HCl gas (gas flow: 40 sccm) at 900 °C to form GaCl precursor. The GaCl was then transported by N$_2$ carrier gas (gas flow: 1600 sccm) into the reaction chamber, where it reacted with NH$_3$ (gas flow: 200 sccm) at 850 °C to grow a GaN buffer layer, the growth time was set for 5 minutes. Afterwards, the temperature of the reaction chamber was raised to 1000 °C, then GaCl (gas flow of HCl: 40 sccm, carrier gas flow: 1600 sccm) reacted with NH$_3$ (gas flow of NH$_3$: 80 sccm, carrier gas flow: 200 sccm) at this temperature to form the GaN epitaxial layer. The growth was carried out under atmospheric pressure.

### GaN micro-LED array growth

To fabricate ordered GaN pyramid arrays, we first patterned the template surface with a close-packed circular-hole mask. A 100 nm Ti film was deposited by e-beam evaporation and subsequently capped with a 200 nm SiO$_2$ layer via plasma-enhanced chemical vapor deposition (PECVD). Standard photolithography and reactive-ion etching were then used to open 3 μm-diameter patterned holes.

Afterwards, GaN micro-LED pyramid arrays were fabricated on patterned mask templates, including four parts: n-GaN, five pairs of MQWs, five pairs of electron blocking layers, and p-GaN were sequentially grown by metal organic chemical vapor deposition (MOCVD). Specifically, a 3 μm-thick n-type GaN:Si layer was grown at 1000°C, followed by the deposition of five periods of InGaN (2.5 nm)/GaN (12.5 nm) MQWs at 760°C, and five periods of AlGaN (2.5 nm)/GaN (12.5 nm) electron blocking layers at 800°C. Finally, a 200 nm-thick p-type GaN:Mg layer was grown at 950°C. The sample was then annealed in a nitrogen atmosphere at 700°C for 30 min to activate the p-GaN. After device fabrication, a 200 nm-thick patterned ITO film was deposited as the top electrode by electron-beam evaporation, and the Ti layer within the template served as a common cathode bottom electrode. For comparison, single-pixel (s-LED) devices were also fabricated on $c$-plane sapphire substrates using the same process.

**Scanning electron microscopy.**

SEM images were obtained using a Field Emission Scanning Electron Microscope (Gemini 500, Zeiss). The samples were primarily imaged using the $SE_2$ and Inlens detectors. The chamber vacuum was maintained at $5\times10^{-10}$ Torr, with an operating voltage of 5 kV and a beam current of 200 pA.

EBSD images were characterized using an Analytical Field Emission Scanning Electron Microscope (Gemini 450, Zeiss). The sample was mounted on a stage tilted at 70°, the chamber vacuum was maintained at $5\times10^{-10}$ Torr, with an operating voltage of 20 kV and a beam current of 10 nA, working distance of 12 mm and stpe size of 0.5 μm.

ECCI images were characterized using a Field Emission Scanning Electron Microscope (Gemini 500, Zeiss). The sample was initially positioned on a stage with 0° tilt. The accelerating voltage was set to 20 kV, the beam current to 3 nA, and the working distance to 5 mm. The targeted diffraction vector $g$ for imaging was [0001].

**Atomic force microscopy.**

An AFM (Dimension ICON, Bruker) was utilized to characterize the surface morphology of films, with non-metal-coated probes (RTESP-300, Bruker) in standard tapping mode.

**X-ray diffraction.**

XRD $2\theta$ scan measurements, $\omega$ scan measurements, and $\varphi$ scan measurements were conducted using a D8 Discover High Resolution Thin-Film X-ray Diffractometer (HRXRD). Wide range $2\theta$ scans were performed with a step size of 0.01° to locate the approximate positions of all relevant diffraction peaks. High-resolution $\omega$ scans (step size: 0.001°) were then conducted around the peaks of interest to accurately determine their peak positions and FWHM. $\varphi$ scans were carried out over a range of 0° to 360° with a step size of 0.5°.

**Two-beam bright/dark-field transmission electron microscope.**

TEM experiments were performed using a Talos F200X G2 Field Emission High resolution TEM (FE-HRTEM). The imaging parameters for the two-beam dark-field and bright-field images are as follows: for the dark-field image, the diffraction vector $g$ was used, with an objective aperture diameter of 10 μm, an accelerating voltage of 200 kV, and an exposure time typically set to 2 s; for the bright-field image, the objective aperture diameter was also 10 μm, the accelerating voltage was 200 kV, and the exposure time was typically set to 0.5 s.

**Scanning transmission electron microscope.**

STEM experiments were carried out using a Spectra Ultra 300KV Double Cs-Corrected TEM. The samples were characterized under 300 kV and 20 pA. A Leica High Vacuum Coating Instrument was used to deposit 15-nm-thick C and Pt layers, serving as conductive and protective coatings, respectively.

TEM and STEM samples were prepared using a Helios 5 UX focused ion beam system. The thinning process was conducted at 30 kV and 0.45 nA, followed by a final cleaning/polishing step at 5 kV and 12 pA.

**Cathodoluminescence measurement**

Spatially resolved cathodoluminescence (CL) measurements were performed using a Hitachi scanning electron microscope (SU8600/Monnac Pro, Hitachi). The CL spectrum was acquired under conditions of 5 kV and 10 μA, with a scan range from 300 to 700 nm, a step size of 1 nm, and a dwell time of 2 s per point.

**Photoluminescence measurement**

Continuous-wave PL spectra measurements were collected using a WITec Alpha300RAS fast imaging Raman microscopy system with a 405 nm wavelength laser. The PL measurement was performed using a 405 nm excitation wavelength with a spot size of approximately 10 μm, under continuous-wave mode. The sample temperature, controlled by a liquid helium cryostat, was initially cooled to 5 K and then varied from 5 K up to room temperature. The spectral scan range was set from 405 nm to 700 nm.

**Fabrication of TLM device**

TLM device fabrication started with processing of n-Gan epitaxial layer on Cu foils. First, mesa isolation area was defined by patterning via maskless lithography (Microwriter ML3, photoresist: AZ1518) with a 600 nm $SiO_2$ PECVD (PD-220NL, Samco) hard mask. Subsequently, mesa isolation was achieved by inductively coupled plasma reactive ion etching (RIE-230iP, Samco) using $SiCl_4$ gas, which extended into the GaN buffer layer. After etching, the $SiO_2$ hard mask was removed by a diluted HF solution (1:10) to expose a clean n-GaN surface. Finally, the ohmic contacts pads were defined by patterning photolithography and metallized by e-beam (ULVAC ei-5z) evaporation of 100nm Ti.

**Contact resistant measurement**

All measurements of contact resistant were conducted in an ambient environment at room temperature. Current-voltage curves characteristics were obtained by a Keysight B1500A Semiconductor Parameter Analyzer with Summit 1100B-M Probe Station.

**Temperature-power measurement**

Temperature-power measurements were conducted in an ambient environment at room temperature. Voltage-time curves were obtained using a B1500A Semiconductor Parameter Analyzer with Summit 1100B-M Probe Station. The surface temperature of the sample was monitored using an infrared thermal imager mounted above the equipment.

**DFT Calculations**

The Cu/AlN interface supercell was constructed to reproduce the experimentally observed epitaxial registry. A Cu (111) slab and an AlN (0001) slab were oriented such that the in-plane $[11\bar{2}0]$ direction of AlN is parallel to the $[1\bar{1}0]$ direction of the Cu (111) surface. The simulation cell dimensions are 15.55 Å, 21.54 Å, and 47.98 Å along the X, Y, and Z directions, respectively, and the supercell contains 240 Al atoms, 240 N atoms, and 420 Cu atoms. First-principles calculations based on DFT were carried out using Vienna ab initio simulation package (VASP) code [72] within the projector augmented wave (PAW) pseudopotential method [73]. Perdew-Burke-Ernzerhof (PBE) form of generalized gradient approximation (GGA) was adopted in the exchange-correlation functional [74]. A plane wave cut-off energy of 500 eV and a Γ-centered k-point mesh 3×3×1 was used. The interfaces were relaxed until the total force of an atomic configuration and the total energy of the system was less than 0.01eV/Å and $10^{-6}$ eV/cell, respectively.

**COMSOL simulation**

The deformation and thermal-mismatch analyses were carried out using COMSOL Multiphysics. The simulations employed a coupled heat transfer in solids and solid mechanics (thermal expansion) multiphysics framework. All models used a multilayer stack with a lateral area of 1 mm × 1 mm as an equivalent representation, with individual layer thicknesses shown in Figures S10a and S10b. Material parameters were taken from the built-in COMSOL material library, including the coefficient of thermal expansion, density, Young's modulus, and Poisson's ratio.

To establish a stress-free reference state (i.e., zero thermal mismatch), the initial substrate temperature was set to 1300 K. The interfacial thermal resistance was not included in the model because its effect was found to be negligible. All external boundaries were assumed to be thermally insulated, and each material was treated linearly elastic and allowed to undergo free thermal expansion within the computational domain.

Mechanical constraints were applied to the substrate (sapphire or W) to suppress rigid-body motion. The meshing strategy used for the multilayer structure is shown in Figure S10, and a high mesh quality was maintained (average mesh quality of 0.5). Based on this configuration, the steady-state residual stress and strain in each layer were calculated after the structure cooled to room temperature (300 K).

## Acknowledgements

This work was supported by the National Natural Science Foundation of China (62574168, 92577201, T2188101, 52192614), and the Research Center for Industries of Future (RCIF, WU2022C022) and Westlake Education Foundation. H.C. and Y.Z. acknowledge the Digital Research Alliance of Canada.The authors thank the Westlake Instrumentation and Service Center for Physical Sciences (ISCPS), Instrumentation and Service Center for Molecular Sciences and Center for Micro/Nano fabrication at Westlake University.

## Author contributions

W. Kong, X. Chen, and K. Liu conceived the idea and planned the research. Y. Ma carried out the HVPE growth and material characterization. J. Yang designed the HVPE systems. P. Jiang performed the PVD sputtering deposition. J. Cao conducted device fabrication and characterization. H. Zhu performed the TEM and strain analyses. Y. Song and T. Wei carried out the MOCVD growth. M. He, M. Wu, and K. Liu provided the single-crystalline Cu foils. X. Xu and T. Jiang performed the COMSOL simulations and analysis. H. Chen, H. He, and Y. Zhou carried out the DFT calculations and analysis. H. Wang performed the CL characterization. All authors contributed to the discussion and analysis of the results regarding the manuscript. W. Kong supervised the project.

## Competing interests

The authors declare no conflict of interest.

## Data availability statement

The data that support the findings of this study are available from the corresponding author upon reasonable request.

## Supplementary information

The online version contains supplementary materials including Fig. S1 – Fig. S15 and Table S1.

58 Liu, Y. *et al.* Ultra-low-defect homoepitaxial micro-LEDs with enhanced efficiency and monochromaticity for high-PPI AR/MR displays. *Photonix* **5**, 23 (2024).

59 Li, Z. *et al.* Vertical GaN-on-GaN micro-LEDs for near-eye displays. *Adv. Sci.* **12**, 202506784 (2025).

60 Feng, F. *et al.* High-power AlGaN deep-ultraviolet micro-light-emitting diode displays for maskless photolithography. *Nat. Photon.* **19**, 101–108 (2025).

61 Nath, P. & Chopra, K. Thermal-conductivity of copper-films. *Thin Solid Films* **20**, 53–62 (1974).

62 Shibata, H. *et al.* High thermal conductivity of gallium nitride crystals grown by HVPE process. *Mater. Trans.* **48**, 2782–2786 (2007).

63 Baek, W. *et al.* Ultra-low-current driven InGaN blue micro light-emitting diodes for electrically efficient and self-heating relaxed microdisplay. *Nat. Commun.* **14**, 1386 (2023).

64 Bai, J. *et al.* A Direct epitaxial approach to achieving ultrasmall and ultrabright InGaN micro light-emitting diodes. *ACS Photonics* **7**, 411–415 (2020).

65 Chen, C., Chen, H., Liao, J., Yu, C. & Wu, M. Fabrication and characterization of active-matrix 960 x 540 blue GaN-based micro-LED display. *IEEE J. Quantum Electron.* **55**, 3300106 (2019).

66 Day, J. *et al.* III-Nitride full-scale high-resolution microdisplays. *Appl. Phys. Lett.* **99**, 031116 (2011).

67 Gong, Z. *et al.* Matrix-addressable micropixellated InGaN light-emitting diodes with uniform emission and increased light output. *IEEE Trans. Electron Devices* **54**, 2650–2658 (2007).

68 Herrnsdorf, J. *et al.* Active-matrix GaN micro light-emitting diode display with unprecedented brightness. *IEEE Trans. Electron Devices* **62**, 1918–1925 (2015).

69 Liu, X. *et al.* High-bandwidth InGaN self-powered detector arrays toward MIMO visible light communication based on micro-LED arrays. *ACS Photonics* **6**, 3186–3195 (2019).

70 Templier, F. GaN-based emissive microdisplays: A very promising technology for compact, ultra-high brightness display systems. *J. Soc. Inf. Display* **24**, 669–675 (2016).

71 Wierer, J. & Tansu, N. III-Nitride micro-LEDs for efficient emissive displays. *Laser Photonics Rev.* **13**, 1900141 (2019).
20

Supporting Information for

# Strain-released epitaxy of GaN enabled by compliant single-crystalline metal foils

Yaqing Ma[1, 2, 9], Junwei Cao[1, 2, 9], Huaze Zhu[2, 9]*, Yijian Song[4, 9], Huicong Chen[7], Menglin He[3], Jun Yang[5], Ping Jiang[5], Tong Jiang[2], Han Chen[1, 2], Xiang Xu[1, 2], Yuqiao Zheng[1], Hao Wang[2], Muhong Wu[6], Yu Zou[7], Xiaochuan Chen[8]*, Tongbo Wei[4]*, Kaihui Liu[6]*, Wei Kong[2]*.

[1] School of Materials Science and Engineering, Zhejiang University, Hangzhou 310027, China

[2] Department of Material Science and Engineering, School of Engineering, Westlake University, Hangzhou 310000, China

[3] Zhongke Crystal Materials (Dongguan) Technology Co., Ltd., Dongguan 523000, China

[4] Research and Development Center for Semiconductor Lighting Technology Institute of Semiconductors, Chinese Academy of Sciences, Beijing 100083, China

[5] Westlake Smoky-mountain Technologies, Hangzhou 310000, China

[6] State Key Laboratory for Mesoscopic Physics, Frontiers Science Centre for Nano-optoelectronics, School of Physics, Peking University, Beijing

[7] Department of Materials Science and Engineering, University of Toronto, Toronto, ON, M5S 3E4, Canada

[8] BOE Technology Group Co., Ltd., Beijing 100176, China

[9] These authors contribute equally: Yaqing Ma, Junwei Cao, Huaze Zhu, Yijian Song.

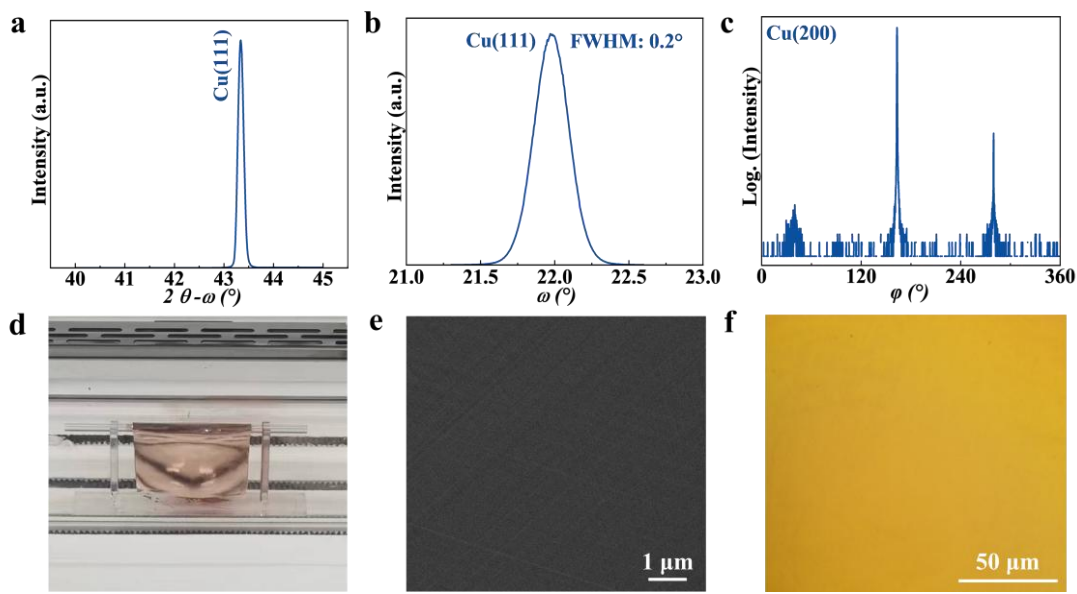

**Fig.S1. Characterization and morphology of the copper (Cu) sample. a,** X-ray diffraction (XRD) 2θ-ω scan of Cu foil (111) crystal plane. **b,** ω scan of Cu (111) plane with a measured full width at half maximum (FWHM) of 0.2°. **c,** φ scan of Cu (200) plane. **d,** Photograph of actual sample. **e-f,** Scanning electron microscopy (SEM) and optical images of surface microstructure of Cu foil.

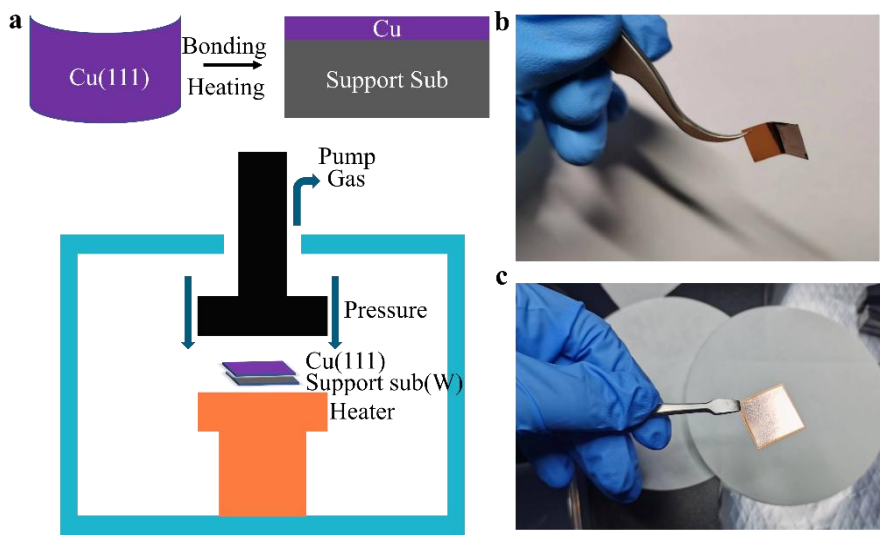

**Fig.S2. Schematic illustration and photographs of the Cu (111) layer bonding process and the resulting samples. a**, Schematic diagram of the bonding process between Cu foil and a tungsten(W) support substrate. **b-c,** Photographs of actual samples before and after bonding process.

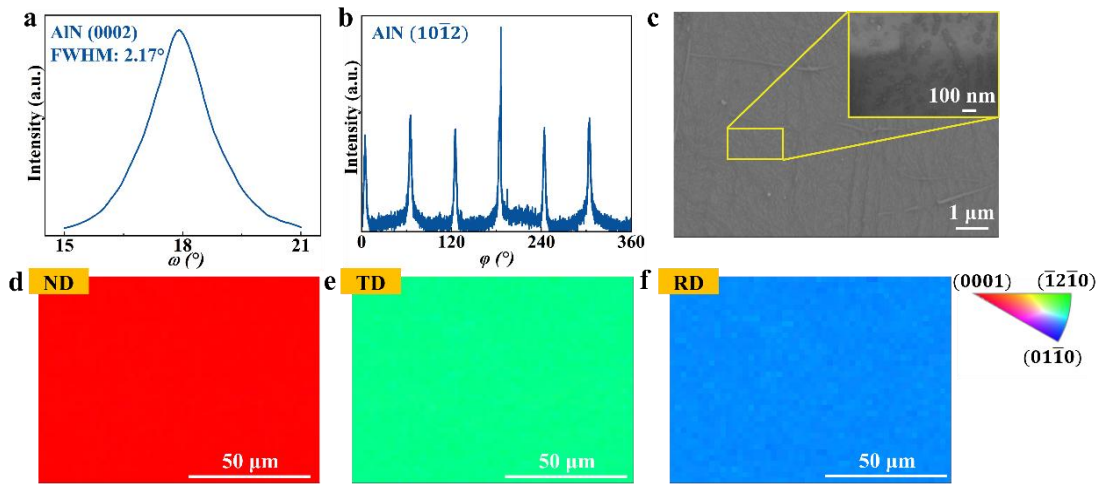

**Fig.S3. Comprehensive structural and microstructural characterization of AlN thin film. a,** XRD $\omega$ scan of AlN (0002) deposited on Cu foils. **b,** XRD $\varphi$ scan of AlN (10$\bar{1}$2) plane. **c,** SEM image of surface morphology of AlN layer on Cu foil. The inner part shows a higher magnification view. **d-f,** Electron backscatter diffraction (EBSD) inverse pole figure maps of the AlN film along the **d** normal direction (ND), **e** transverse direction (TD), and **f** rolling direction (RD), respectively.

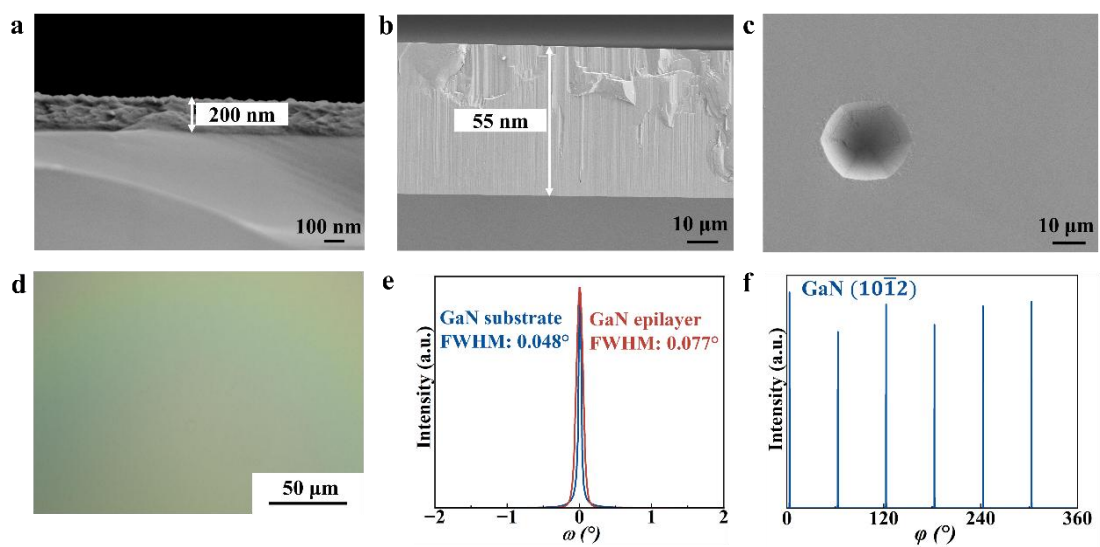

**Fig.S4. Comprehensive microstructural and crystallographic characterization of GaN grown on the engineered Cu substrate. a-b,** Cross-sectional SEM image of cross-sectional structure of GaN low-temperature buffer layer and GaN high-temperature epilayer grown on sapphire substrates, respectively. **c,** SEM image of GaN sample surface. **d,** Optical image of large-area, uniform region of GaN epilayer. **e,** XRD $\omega$ scan comparison of the crystal quality of standard GaN substrate (blue curve, FWHM = 0.048°) and the GaN epilayer grown on sapphire substrate (red curve, FWHM = 0.077°). **f,** XRD $\varphi$ scan of the asymmetrical GaN $(10\bar{1}2)$ plane.

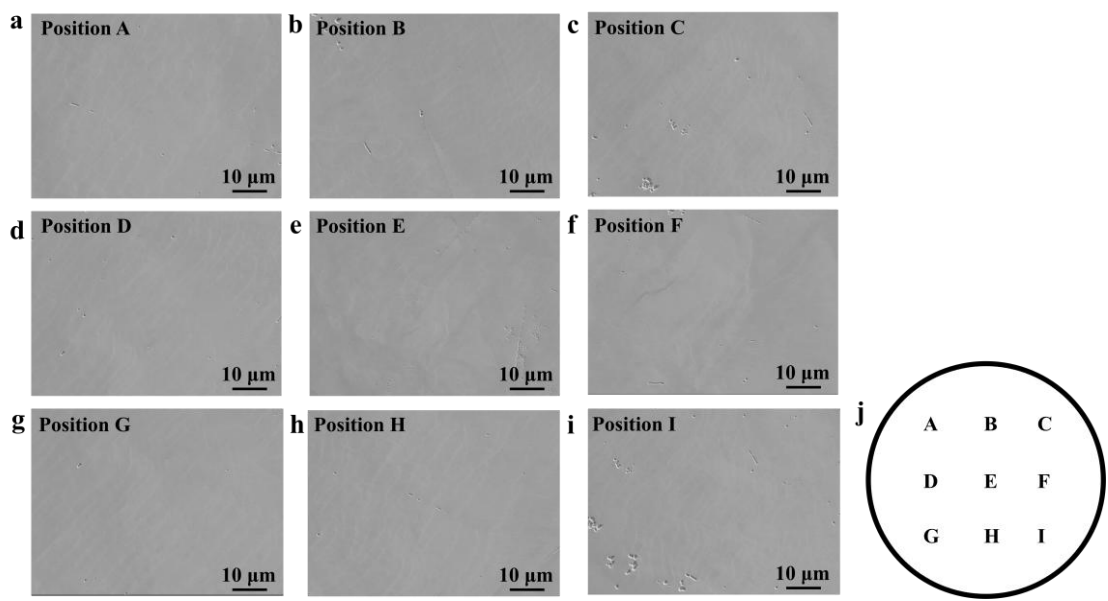

**Fig.S5. Surface SEM across nine representative positions. a-i,** SEM morphology images of the sample surface captured at nine distinct positions, corresponding to the locations labeled in the schematic diagram **j**.

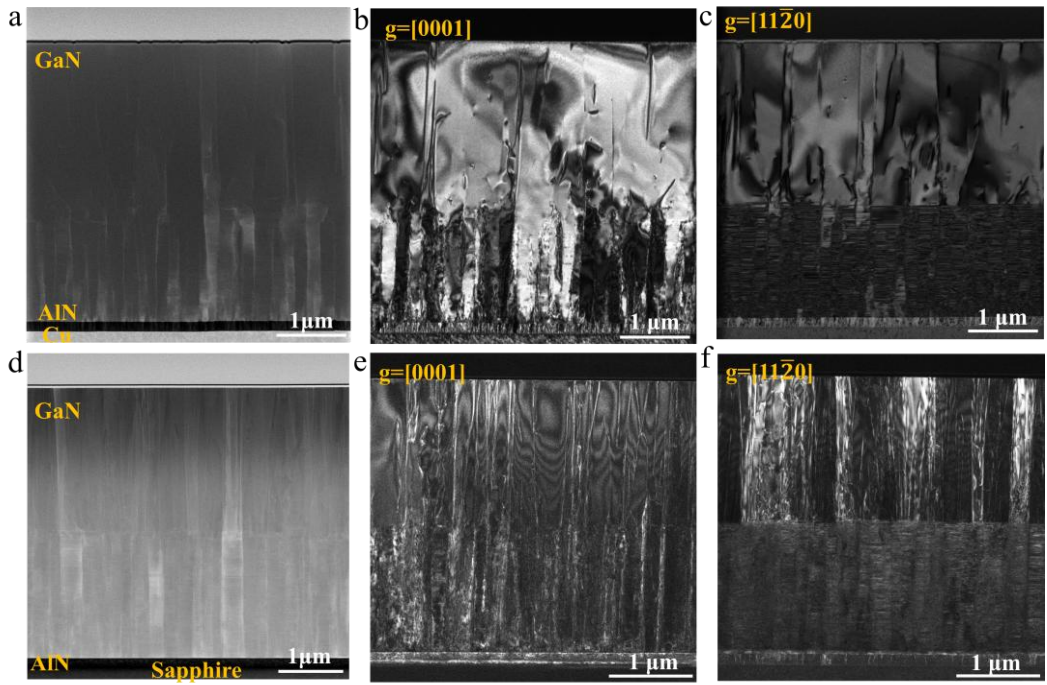

**Fig.S6. Cross-sectional transmission electron microscopy (TEM) characterization of GaN epitaxial layer on different substrates. a,** Cross-sectional TEM image of GaN grown on Cu foil with an AlN buffer layer. **b-c,** Dark-field TEM images of GaN-on-Cu at two-beam condition using diffraction vectors $g$ = [0001] and $g$ = [11$\bar{2}$0], respectively. **d,** Cross-sectional TEM image of GaN grown on a sapphire substrate. **e-f,** Corresponding dark-field images obtained with $g$ = [0001] and $g$ = [11$\bar{2}$0], respectively.

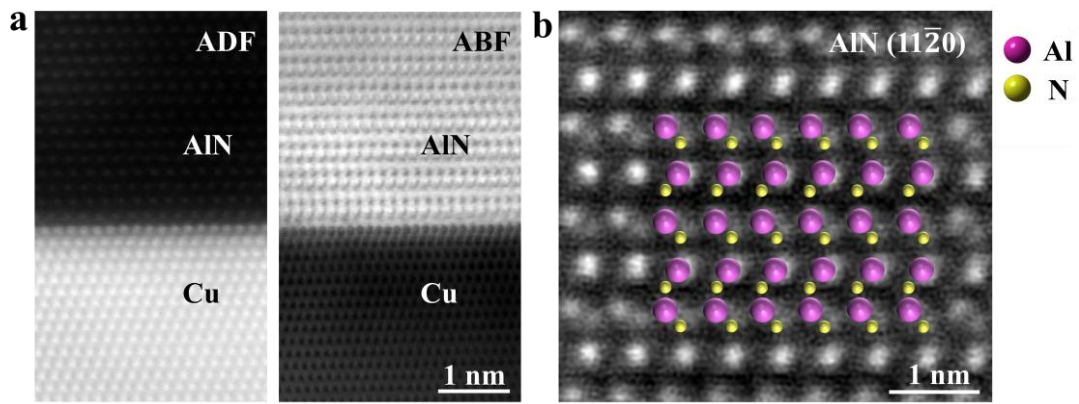

**Fig.S7. Atomic-scale characterization of the AlN/Cu interface structure. a,** Cross-sectional annular dark-field (ADF) and annular bright-field (ABF) scanning transmission electron microscopy (STEM) images of the AlN/Cu interface. **b,** High-angle annular dark-field (HADDF) STEM image of the AlN viewed along the [11$\bar{2}$0] zone axis, overlaid with the corresponding atomic structural model.

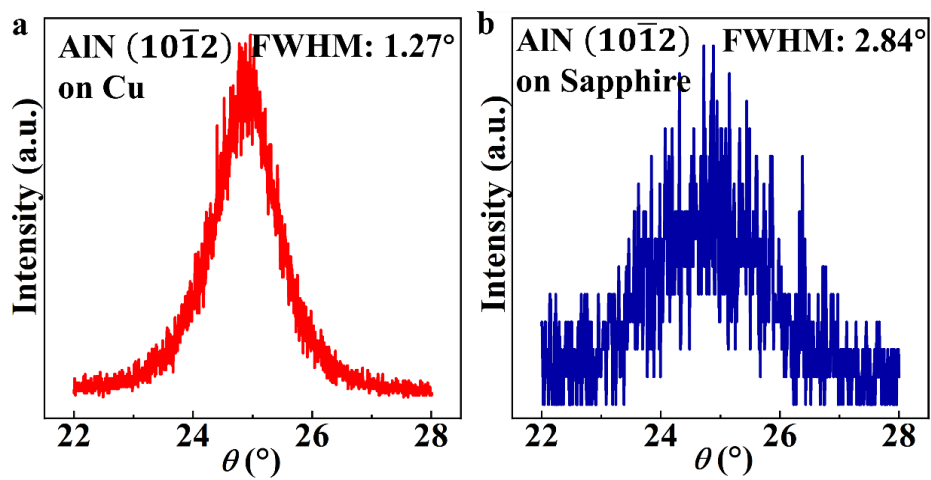

**Fig.S8. Comparison XRD analysis of AlN crystallinity grown on different substrates. a,** XRD $\omega$ scan of the AlN $(11\bar{2}0)$ plane grown on a Cu substrate, showing a FWHM of 1.27°. **b,** XRD $\omega$ scan of the AlN $(11\bar{2}0)$ plane grown on a sapphire substrate, showing a FWHM of 2.84°.

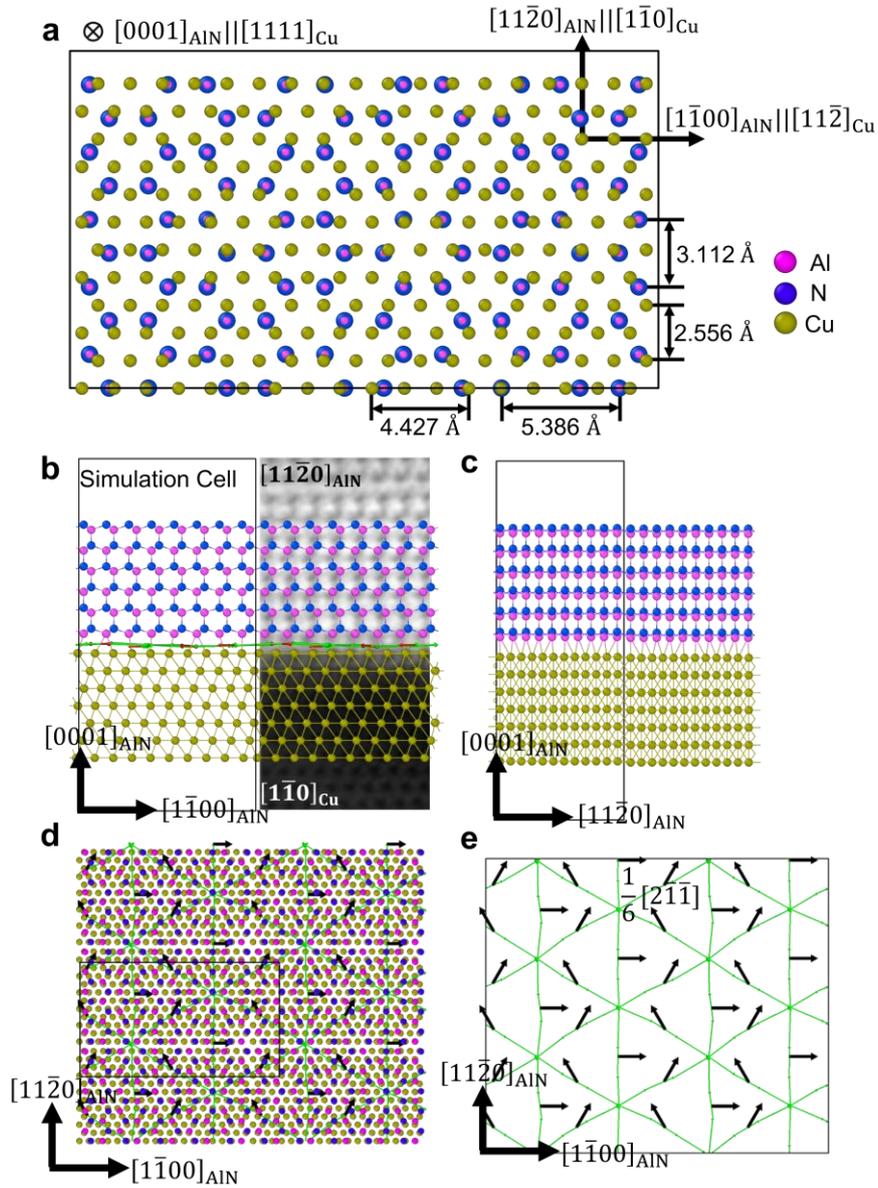

**Fig.S9. DFT model of the AlN/Cu interface**. **a,** In-plane lattice alignment between epitaxial wurtzite AlN and the Cu substrate used to construct the interfacial supercell (Al, N and Cu atoms are indicated by colour). Representative interatomic spacings used to define the registry are labelled. **b,** Atomic model of the AlN/Cu interface within the simulation cell, shown together with an embedded experimental TEM image for comparison. **c,** Cross-sectional view of the interfacial atomic structure corresponding to the left-side view in b. **d,** Plan-view atomic configuration of the interface

corresponding to the top view in b, highlighting the in-plane periodicity relevant to misfit accommodation. **e,** Schematic of the initial two-dimensional interfacial dislocation network; dislocation lines are in green color and the associated Burgers-vectors are highlighted by black arrows.

Using the experimentally observed epitaxial orientation relationship (Fig. S9a), we constructed an atomistic Cu/AlN interface in a periodic slab geometry. The computed equilibrium lattice parameter for fcc Cu (3.61 Å) and Wurtzite AlN (a = 3.11 Å and c = 4.98 Å) are consistent with the values measured in this study. The interfacial supercell was constructed to reproduce the experimentally observed registry by combining a Cu (111) slab and an AlN (0001) slab, with the in-plane $[11\bar{2}0]$ direction aligned parallel to the $[1\bar{1}0]$ on the Cu (111) surface. Along this matching direction, the repeat distances are 3.112 Å for AlN and 2.556 Å for Cu (111) (Fig. S9a), yielding a near-commensurate coincidence lattice in which five AlN repeats approximately match to six Cu repeats (5 × 3.112 Å ≈ 6 × 2.556 Å). This construction reduces the residual in-plane mismatch to a small value (on the order of ~1-2%) distributed over the coincidence period, enabling a realistic description of misfit accommodation at the interface. The resulting interface model is shown in Fig. S9b and c (with representative TEM contrast embedded for comparison) and the corresponding plan-view configuration (Fig. S9d) highlights the periodic interfacial modulation associated with mismatch accommodation. In the initial, unrelaxed structure, the mismatch is naturally expressed as a periodic array/network of misfit dislocations at the interface (Fig. S9e), consistent with the coincidence-period argument above. We then obtained a stable interfacial configuration by fully relaxing the atomic coordinates while holding the simulation cell fixed (Fig. S10a), so that the relaxation reflects local interfacial reconstruction.

The relaxed structure and its derived fields (Fig. S10) indicate that mismatch accommodation is dominated by atomic-scale slip and shear within the Cu side adjacent to the interface. The displacement map (Fig. S10b) shows that relaxation is concentrated near the interface of Cu, whereas the AlN sublattice remains comparatively coherent away from the immediate interfacial region. Consistently, the spatial distribution of the in-plane elastic strain component (Fig. S10c) is localized around the interface of Cu, confirming that the mismatch strain is largely screened within a confined interfacial zone rather than propagating deeply into the AlN film. The corresponding relaxed dislocation network representation (Fig. S10d) provides a compact description of the interfacial misfit accommodation in terms of dislocation line directions and Burgers-vector directions in the AlN basal plane. Together, these results support the experimental inference that the high ductility of the Cu substrate enables strain relief through interfacial slip within Cu atomic planes, thereby suppressing the formation of a high density of misfit dislocations in the epitaxial AlN layer.

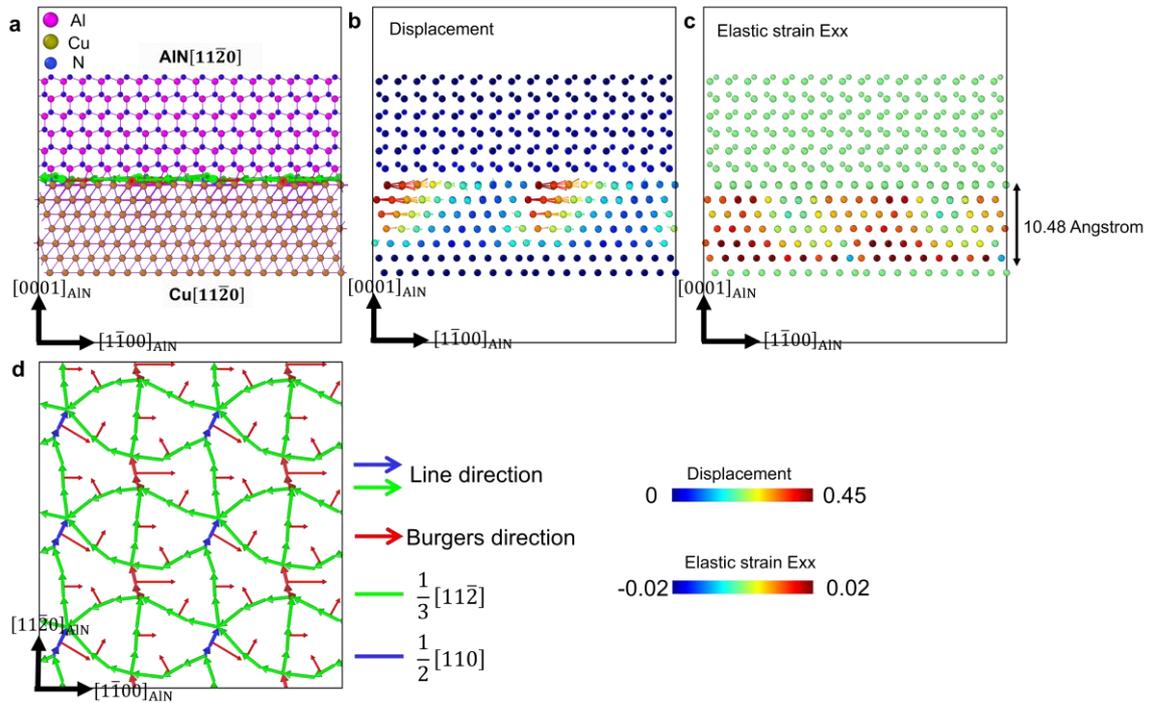

**Fig.S10. Relaxed interfacial structure, displacement field and elastic strain obtained from DFT calculation. a,** Relaxed atomic configuration of the AlN/Cu interface after structural optimisation. **b,** Per-atom displacement magnitude relative to the initial (unrelaxed) configuration, illustrating relaxation concentrated near the interface. **c,** Spatial distribution of the in-plane elastic strain component after relaxation; the marked length indicates the characteristic extent used for strain evaluation. **d,** Interfacial dislocation network in the relaxed state, showing dislocation line directions and corresponding Burgers-vector directions. Atoms in b and c are coloured by displacement magnitude and elastic strain $\varepsilon_{xx}$, respectively.

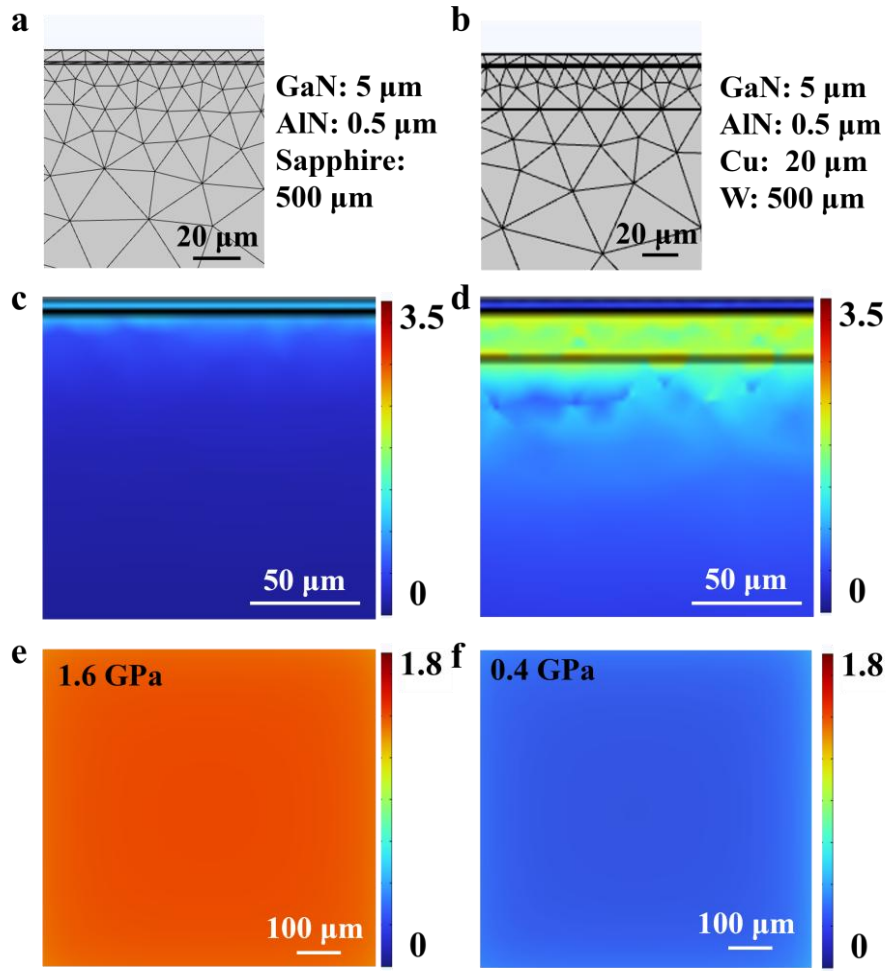

**Fig.S11. Simulation comparison of stress distribution in GaN/AlN heterostructures on different substrate stacks. a,** Schematic diagram of the cross-sectional mesh structure: a 5-μm GaN layer on a 0.5-μm AlN seed layer, directly grown on a 500-μm sapphire substrate. **b,** Schematic diagram of the cross-sectional mesh structure: the same GaN/AlN stack is now grown on a 20-μm Cu foil interlayer, which is bonded to a 500-μm W carrier substrate. **c,** Cross-sectional simulated stress distribution of the GaN-sapphire structure. **d,** Cross-sectional simulated stress distribution of the GaN-Cu-W structure. **e,** Surface stress distribution of GaN-sapphire structure. **f,** Surface stress distribution of the GaN-Cu-W structure.

The deformation and thermal-mismatch analysis were carried out using COMSOL Multiphysics. The simulations employed a coupled heat transfer in solids and solid mechanics (thermal expansion) multi physics framework. All models used a multilayer stack with a lateral area of 1 mm × 1 mm as an equivalent representation, with individual layer thicknesses shown in Figures S10a and S10b. Material parameters were taken from the built-in COMSOL material library, including the coefficient of thermal expansion, density, Young's modulus, and Poisson's ratio.

To establish a stress-free reference state (i.e., zero thermal mismatch), the initial substrate temperature was set to 1300 K. The interfacial thermal resistance was not included in the model because its effect was found to be negligible. All external boundaries were assumed to be thermally insulated, and each material was treated as linearly elastic and allowed to undergo free thermal expansion within the computational domain.

Mechanical constraints were applied to the substrate (sapphire or W) to suppress rigid-body motion. The meshing strategy used for the multilayer structure is shown in Figure S10, and a high mesh quality was maintained (average mesh quality of 0.5). Based on this configuration, the steady-state residual stress and strain in each layer were calculated after the structure cooled to room temperature (300 K).

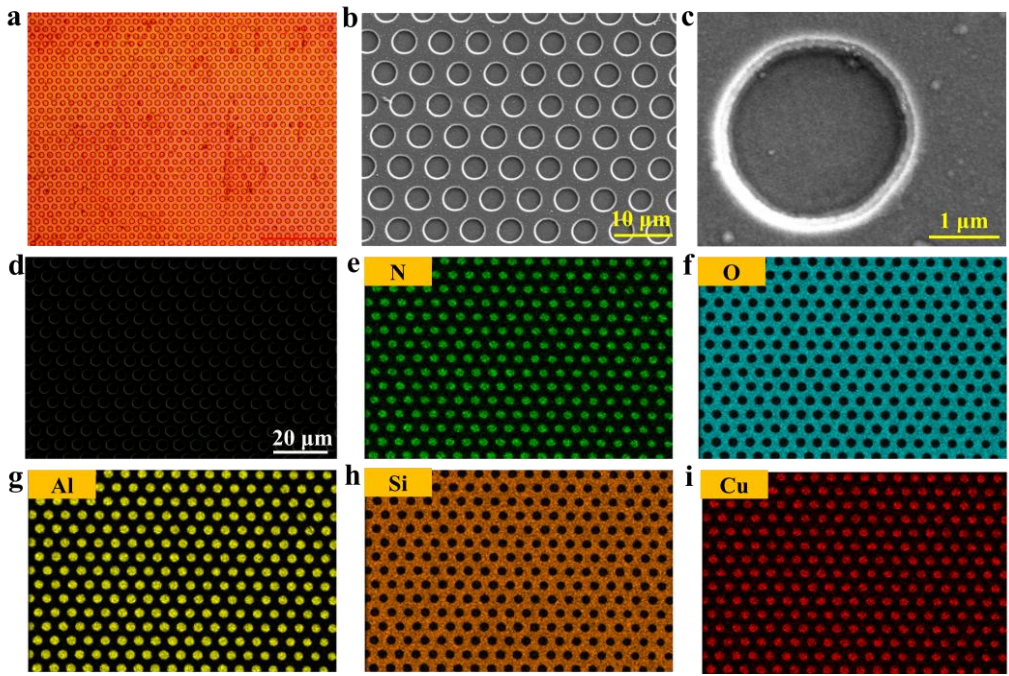

**Fig.S12. Microstructural and elemental characterization of the patterned array. a,** Optical image of highly ordered patterned holes fabricated via lithography. **b,** SEM images of patterned holes with an average $d_{hole}$ value of 3 μm. **c,** SEM image of a single dominant hole with measured size of 2.89 μm. **d-i,** Energy-dispersive X-ray spectroscopy (EDS) elemental maps of the same region, showing the spatial distribution of nitrogen (N, green), oxygen (O, cyan), aluminum (Al, yellow), silicon (Si, orange-brown), and Cu (red).

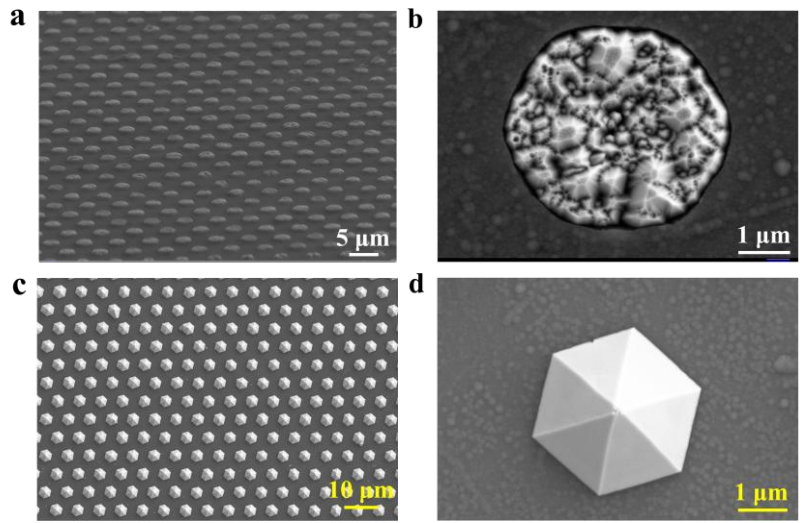

**Fig.S13. Structural evolution of selective-area epitaxy (SAE) of GaN pyramid arrays. a,** SEM image of as-grown LT GaN buffers on highly ordered arrays on AlN/Cu foil substrates. **b,** SEM image of morphology of one dominant patterned GaN buffer. **c,** SEM image of as-grown GaN patterned pyramids on patterned circular holes on AlN/Cu foil substrates. **d,** SEM image of morphology of a single GaN crystal.

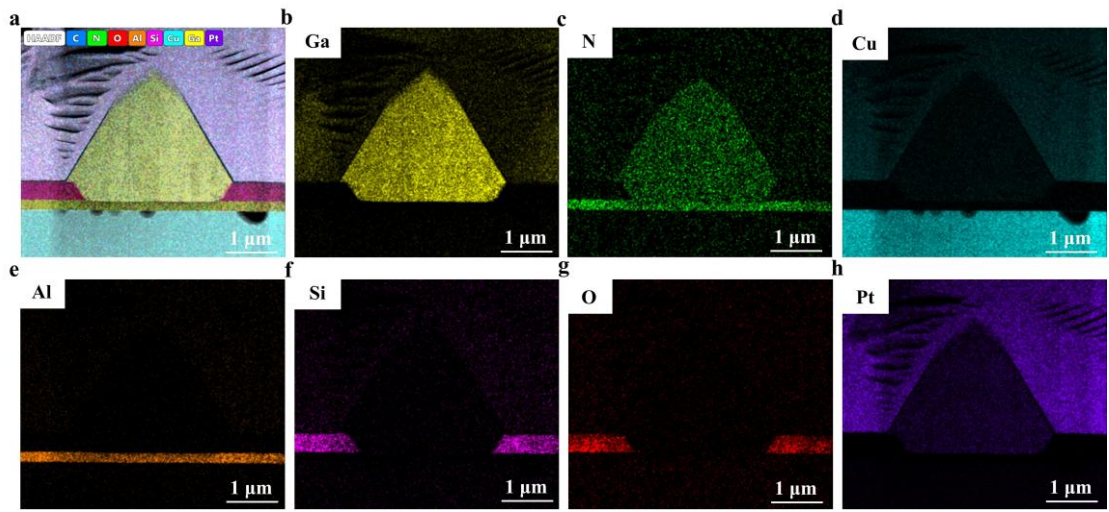

**Fig.S14. EDS mapping of the GaN pyramid heterostructure. a,** Color overlay image combining signals from multiple elements. **b-h,** Individual EDS elemental maps for gallium (Ga, yellow), N (green), Cu (cyan), Al (orange), Si (pink), O (red), and platinum (Pt, blue) elements.

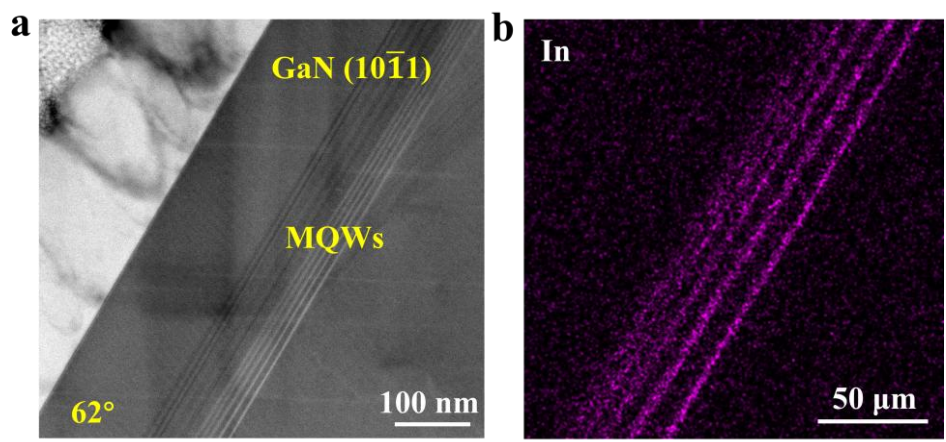

**Fig.S15. Microstructural and compositional analysis of the InGaN/GaN multi-quantum well (MQW) region. a,** Cross-sectional TEM image viewed along a specific zone axis, showing an inclined interface at approximately 62° to the growth plane. The image highlights the periodic layered structure of the MQWs grown on the Ga $(10\bar{1}1)$ facet. **b,** Corresponding SEM-EDS elemental map for indium (In, shown in pink/magenta).

**Table 1. Comparison of crystalline quality of GaN epitaxial layers on different substrates (Excluding selective area growth)**

| Substrate | TDDs (cm$^{-2}$) | Thickness (μm) | Ref. |
|---|---|---|---|
| GaN-Sap | $6.6 \times 10^8$ | ~5 | 1 |
| | ~$1 \times 10^{10}$ | ~5 | 2 |
| | $7 \times 10^8$ | ~5 | 3 |
| | $5 \times 10^9$ | ~5 | 4 |
| | $1 \times 10^8 - 1 \times 10^{10}$ | ~5 | 5 |
| GaN-SiC | $4 \times 10^8$ | 3.5 | 6 |
| | $7 \times 10^7$ | ~1 | 7 |
| | $5.5 \times 10^8$ | ~5 | 8 |
| GaN-Si | ~$1 \times 10^{10}$ | ~5 | 2 |
| | ~$10^9$ | ~5 | 9 |
| | $2.5 \times 10^9$ | 2.5 | 10 |
| | $5.8 \times 10^9$ | 2 | 11 |
| | $5 \times 10^9$ | 6.3 | 12 |
| | $1.1 \times 10^9$ | 6.7 | 13 |